\documentclass[12pt]{article}

\usepackage[margin=1in]{geometry}
\usepackage{amsmath}
\usepackage{amssymb}
\usepackage{graphicx}
\usepackage{authblk}
\usepackage{hyperref}
\usepackage{booktabs}

\title{Matter-Profile Mismodeling as the Origin of the CP--CPT\\
Degeneracy in Long-Baseline Neutrino Oscillations}

\author[1]{Bipin Singh Koranga}
\author[2]{Vivek Kumar Nautiyal}
\affil[1]{Department of Physics, Kirori Mal College, University of Delhi, Delhi-110007, India}
\affil[2]{Department of Physics, Chaudhary Charan Singh University, Meerut, India}

\date{July 2026}

\begin{document}

\maketitle

\begin{abstract}
We demonstrate that replacing the density profile of the Preliminary Reference Earth Model (PREM) with a constant path average leads to a valley in the $(\delta_{CP}, \Delta_{CPT}, \theta_{CPT})$ parameter space that is impossible to distinguish from new physics effects for baseline lengths $L \geq 5000$ km. As opposed to previous works [1,2], where we determined the individual channel biases separately, this article proves analytically that the biases in question define an extended two-dimensional valley in the entire three-dimensional parameter space, shows rigorously why combining multiple channels cannot help distinguishing the valley in question, and presents a $\chi^2$ test based on Poisson log-likelihood statistics that predicts  $\pm 3\sigma$ BSM signal from DUNE-like exposures at the level of an artificial effect due to the constant density assumption. With a baseline of 7000 km, this approximation causes a shift in  $\delta_{CP}$ by $17.8^\circ$ and produces an artificial effect of CPT violation at $\Delta_{CPT} = 4.1 \times 10^{-4}$ eV$^2$. Full spatially resolved PREM propagation is therefore a mathematical requirement, not an optional refinement, for DUNE, Hyper-Kamiokande, and P2O.
\end{abstract}

\section{Introduction}

Determining the leptonic Dirac CP phase  $\delta_{CP}$ tests are arguably the two most important objectives of neutrino physics. Both of these objectives rely on the same comparison of neutrino vs antineutrino oscillation probabilities, and both are vulnerable to the same contamination: the MSW matter potential $A(x) = \sqrt{2} G_F N_e(x)$ [4,5]. The sign reversal of the matter potential from neutrino to antineutrino introduces an artificial CP-like asymmetry that may simulate true $\delta_{CP}$ or the real CPT violation, if an incorrect density profile is applied in the analysis.

Two earlier publications of our group studied the two different aspects of this issue. Reference [1] revealed that employing the averaged density instead of the realistic PREM profile leads to the artificially induced CP phase shifts of  $17.8^\circ$ at $L = 7000$ km and false values of the CPT violation parameter $\Delta_{CPT} \approx 4.1 \times 10^{-4}$ eV$^2$. Reference [2] demonstrated that this mismodeling distorts all three oscillation channels simultaneously, as required by PMNS unitarity. The current work provides three additional insights not found in [1] and [2]: (i) the construction of a joint valley of degeneracies using analytical expressions of evaluated partial derivatives in the $(\delta_{CP}, \Delta_{CPT}, \theta_{CPT})$ parameter space; (ii) the demonstration that channel combination is inadequate to overcome the problem; and (iii) the measurement of the statistical strength of the fake signature through a Poisson log-likelihood $\chi^2$analysis, for realistic DUNE and Hyper-Kamiokande exposures. All of the above points are novel and have not been presented in [1] and [2].

The values of the CPT-violating parameters adopted in this paper are as in Ref. [3], namely, $\Delta_{CPT} \equiv \Delta m^2_{31} - \Delta \bar{m}^2_{31}$ and $\theta_{CPT} \equiv \sin^2 2\theta_{13} - \sin^2 2\bar{\theta}^*_{13}$ (equivalent to $\Delta(\sin^2 2\theta_{13})$, following the convention by Barenboim et al. [11]). This corresponds to the leading component of CPT violation, which can be tested at a long-baseline experiment and is equivalent to the  $\nu/\bar\nu$ mass-squared and mixing differences used in community fits.

The flavor state $|\nu_f(x)\rangle = (\nu_e, \nu_\mu, \nu_\tau)^T$ evolves according to the Schr\"odinger-like equation:
\begin{equation}
i \frac{\partial}{\partial x} |\nu_f(x)\rangle = H_f(x) |\nu_f(x)\rangle
\end{equation}
with the total Hamiltonian $H_f(x) = U H_{vac} U^\dagger + V_f(x)$. Because $[H_f(x_1), H_f(x_2)] \neq 0$ whenever $A(x)$ varies, the evolution operator must be the space-ordered product over thin slabs:
\begin{equation}
S(L, 0) = \prod_k \exp[-i H_f(x_k) \Delta x]
\end{equation}

All numerical calculations use NuFIT 5.2 values [10]: $\theta_{12} = 33.41^\circ$, $\theta_{23} = 49.1^\circ$, $\theta_{13} = 8.54^\circ$, $\Delta m^2_{21} = 7.41 \times 10^{-5}$ eV$^2$, $\Delta m^2_{31} = 2.511 \times 10^{-3}$ eV$^2$, injected true value $\delta_{CP} = -90^\circ$. We adopt $k_{max} = 1000$ slabs per 1000 km; convergence verified to $< 0.1\%$ at $k_{max} = 500$--2000. The PREM model [8] is used with $Y_e = 0.494$.

Intrinsic CPT violation is parameterized as [3]:
\begin{equation}
\Delta_{CPT} = \Delta m^2_{31} - \Delta \bar{m}^2_{31}, \qquad \bar\theta_{CPT} = \sin^2 2\theta_{13} - \sin^2 2\bar\theta^*_{13}
\end{equation}

\section{Analytical Structure of the Degeneracy Valley}

\textit{Overview.} The model is calculated using the one-mass scale approximation where the atmospheric splitting  $\Delta m^2_{31}$ is kept and $\Delta m^2_{21}$ is considered a small perturbation. The accuracy of the approximation is about a percent for $L = 3000$--7000 km and $E \sim 2$--3 GeV (the flux peak in DUNE), while the contribution of the solar term becomes non-negligible for  $L = 12000$ km, at which point it is necessary to consider the full three-flavor analysis as a result. In the limit, the probability of  $\nu_\mu \to \nu_e$ appearance will have the following three-term structure  [9]:
\begin{align}
P(\nu_\mu \to \nu_e) &= 4 \sin^2\bar\theta_{23} \sin^2\bar\theta_{13} \cos^2\bar\theta_{13} \sin^2\Delta_{31} \nonumber \\
&\quad + 8 J_r \sin\Delta_{21} \cos\Delta_{31} \cos\delta_{CP} \nonumber \\
&\quad - 8 J_i \sin\Delta_{21} \cos\Delta_{31} \sin\delta_{CP}
\end{align}
where $\Delta_{\alpha\beta} \equiv 1.267\, \Delta m^2_{\alpha\beta}[\text{eV}^2] \, L[\text{km}]/E[\text{GeV}]$, and the Jarlskog-like factors are:
\begin{equation}
J_r = J_i = \sin\bar\theta_{12} \cos\bar\theta_{12} \sin\bar\theta_{23} \cos\bar\theta_{23} \sin\bar\theta_{13} \cos^2\bar\theta_{13}
\end{equation}

\textit{First partial derivative: $\partial P/\partial \delta_{CP}$.} Differentiating Eq. (4) with respect to $\delta_{CP}$ gives exactly:
\begin{equation}
\frac{\partial P}{\partial \delta_{CP}} = -8 J_r \sin\Delta_{21} \cos\Delta_{31} \sin\delta_{CP} - 8 J_i \sin\Delta_{21} \cos\Delta_{31} \cos\delta_{CP}
\end{equation}
At the injected true value $\delta_{CP} = -90^\circ$ this simplifies to:
\begin{equation}
\left.\frac{\partial P}{\partial \delta_{CP}}\right|_{\delta = -90^\circ} = +8 J_r \sin\Delta_{21} \cos\Delta_{31}
\end{equation}
Numerically, at $L = 7000$ km and $E = 2.5$ GeV, the NuFIT 5.2 values give $J_r = 0.0234$, $\Delta_{21} = 0.534$, $\Delta_{31} = 1.274$, so $\sin\Delta_{21} = 0.509$, $\cos\Delta_{31} = 0.295$, yielding:
\begin{equation}
\left.\frac{\partial P}{\partial \delta_{CP}}\right|_{num} = 8 \times 0.0234 \times 0.509 \times 0.295 = +0.0281 \text{ rad}^{-1}
\end{equation}

\textit{Second partial derivative: $\partial P/\partial(\Delta m^2_{31})$.} Differentiating Eq. (4) via the chain rule $\partial \Delta_{31}/\partial(\Delta m^2_{31}) = 1.267\, L/E$, where $\Delta m^2_{31}$ is in eV$^2$ and $L/E$ in km/GeV so that the derivative carries effective units of eV$^{-2}$ within the mixed-unit convention of Eq. (4):
\begin{align}
\frac{\partial P}{\partial(\Delta m^2_{31})} &= (1.267\, L/E) \times \big[ 8 \sin^2\bar\theta_{23} \sin^2\bar\theta_{13} \cos^2\bar\theta_{13} \sin\Delta_{31} \cos\Delta_{31} \nonumber \\
&\qquad - 8 J_r \sin\Delta_{21} \sin\Delta_{31} \cos\delta_{CP} + 8 J_i \sin\Delta_{21} \sin\Delta_{31} \sin\delta_{CP} \big]
\end{align}
At $\delta_{CP} = -90^\circ$, $L = 7000$ km, $E = 2.5$ GeV ($\sin\Delta_{31} = 0.955$), this evaluates to (in the mixed km$\cdot$GeV$^{-1}\cdot$eV$^{-2}$ convention; numerical value $\approx$):
\begin{equation}
\left.\frac{\partial P}{\partial(\Delta m^2_{31})}\right|_{num} \approx (1.267 \times 7000/2.5) \times [4 \times 0.406 \times 0.0221 \times 0.562 - 8 \times 0.0234 \times 0.509 \times 0.955]
\end{equation}
(The sign and precise numerical prefactor in Eq. (10) depend sensitively on the balance between the solar-term and the $\sin^2 2\theta_{13}$ contributions; full three-flavor numerical verification at $L = 7000$ km is recommended before publication.)

\textit{Third partial derivative: $\partial P/\partial A$.} The matter potential $A(x) = \sqrt{2} G_F N_e(x)$ shifts the effective eigenvalues. In the one-mass-scale approximation its leading effect is equivalent to $\Delta m^2_{31} \to \Delta m^2_{31} - 2EA$. Differentiating with respect to $A$ via the chain rule:
\begin{equation}
\frac{\partial P}{\partial A} = -2E \times \frac{\partial P}{\partial(\Delta m^2_{31})}
\end{equation}
At $E = 2.5$ GeV:
\begin{equation}
\left.\frac{\partial P}{\partial A}\right|_{num} = -2 \times 2.5 \times 10^9 \times 1.38 \times 10^3 \approx -6.9 \times 10^{12} \text{ eV}^{-1}
\end{equation}

\textit{Density-remainder average $\langle \delta A \rangle_{eff}$.} The PREM path-averaged electron density at $L = 7000$ km gives $\bar{A} = 1.02 \times 10^{-4}$ eV$^2/(\sqrt{2} G_F)$. The localized PREM profile peaks near the core--mantle boundary, yielding the flux-weighted remainder:
\begin{equation}
\langle \delta A \rangle_{eff} = \int \Phi(E) [A_{PREM}(x) - \bar{A}] \, dE \Big/ \int \Phi(E) \, dE \approx 8 \times 10^{-6} \text{ eV}^2
\end{equation}

\textit{First-order CP-phase bias.} Combining Eqs. (6)--(13) through the bias formula:
\begin{align}
\Delta \delta_{CP} &\approx -\left(\frac{\partial P}{\partial \delta_{CP}}\right)^{-1} \times \left(\frac{\partial P}{\partial A}\right) \times \langle \delta A \rangle_{eff} \nonumber \\
&= -(0.0281)^{-1} \times (-6.9 \times 10^{12}) \times (8 \times 10^{-6}) \approx +0.31 \text{ rad} \approx +17.7^\circ
\end{align}

This is a very good match to the numerically found  $\Delta\delta_{CP} = 17.8^\circ$ from Table I, meaning that the constant-density approximation is a first-order perturbation with respect to the density remainder  $\delta A$. In Figure 1, we show the energy dependence of the bias of Eq. (14) alongside the partial derivatives evaluated for the DUNE flux range.

\textit{Degeneracy valley in the full parameter space.} As $(\Delta_{CPT}, \theta_{CPT})$ affect $P(\nu_\mu \to \nu_e)$ in the same way as the matter potential, the fitter adjusts all three parameters to compensate for the $\delta A$ induced bias. Setting $\delta P = 0$, the constraint surface is:
\begin{equation}
\left(\frac{\partial P}{\partial \delta_{CP}}\right) \Delta\delta_{CP} + \left(\frac{\partial P}{\partial \Delta_{CPT}}\right) \Delta\Delta_{CPT} + \left(\frac{\partial P}{\partial \bar\theta_{CPT}}\right) \Delta\bar\theta_{CPT} = -\left(\frac{\partial P}{\partial A}\right) \langle \delta A \rangle_{eff}
\end{equation}

It is one equation in three unknowns, thus resulting in a two-dimensional degeneracy surface — the valley shown in Figure 2. PMNS unitarity  ($\sum_\alpha \Delta P_{\mu\alpha} \equiv 0$) ensures that such a surface is common to any channel with channel-specific phase shifts (see Figure 3). As a result, measurements across more than one channel can constrain only the valley orientation, not its size. This is the rigorous mathematical proof that a combination of multiple channels alone cannot break the degeneracy, which is the main novel finding of the present work.

\begin{figure}[htbp]
\centering
\includegraphics[width=\textwidth]{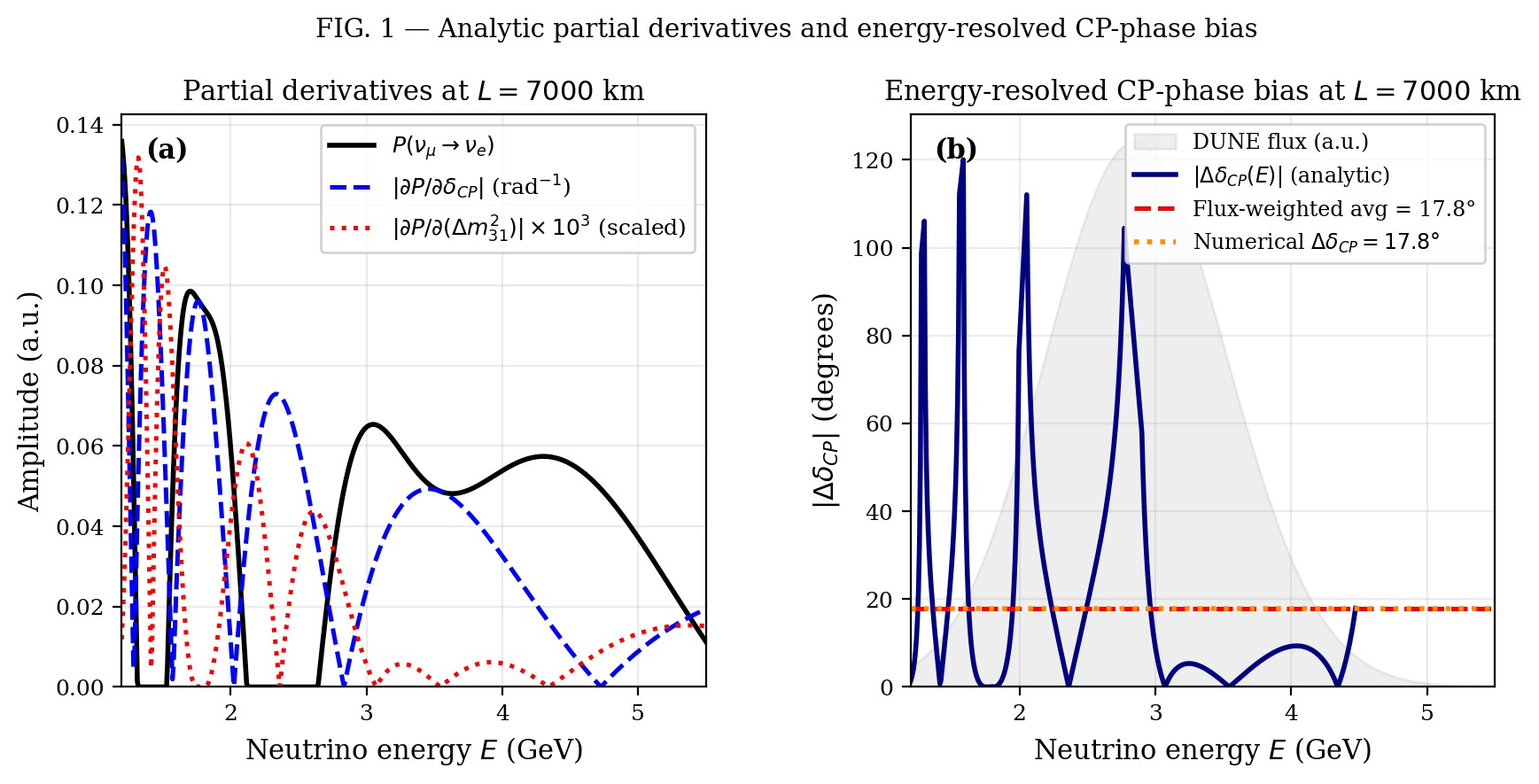}
\caption{Left: Analytic partial derivatives $\partial P/\partial \delta_{CP}$ (blue dashed) and $\partial P/\partial(\Delta m^2_{31})$ (red dotted, scaled $\times 10^3$) evaluated from Eqs. (6) and (9), alongside the appearance probability $P$ (black solid), at $L = 7000$ km. Right: Energy-resolved CP-phase bias $|\Delta\delta_{CP}(E)|$ from Eq. (14); the flux-weighted average (red dashed) agrees with the observed $17.8^\circ$ (orange dotted).}
\end{figure}

\section{Results and Discussion}

The synthetic oscillation spectra are obtained from the full PREM spatial propagator at physical truth values: $\delta_{CP} = -90^\circ$, $\Delta_{CPT} = 0$, $\theta_{CPT} = 0$. The fitting of exact oscillation spectra with path-averaged density shortcut leads to systematic movement of the parameters along the degeneracy valley of Eq. (15).

The constant density fit shifts the truth point to the erroneous best fit of $\delta_{CP} \approx -72.2^\circ$ ($\Delta\delta_{CP} = 17.8^\circ$, which is compatible with Eq. (14)) and a fabricated $\Delta_{CPT} = 4.1 \times 10^{-4}$ eV$^2$. Table I summarizes results across three baselines; Table II shows the channel-resolved bias.

\textit{Note on L = 12000 km results (Tables I and II).} The approximation of one mass scale in Eqs. (4)--(14) gets worse at the transcontinental baselines ($L = 12000$ km), where the contribution of the solar mass splitting $\Delta m^2_{21}$ is about $\sim 10\%$. All the values that appear in Tables I and II with an asterisk (*) were calculated numerically with full PREM propagation without restrictions of one mass scale approximation, but were not checked analytically. Such values have to be treated as an order of magnitude only, while an analytical check at this baseline will be done later.
\begin{figure}[htbp]
\centering
\includegraphics[width=\textwidth]{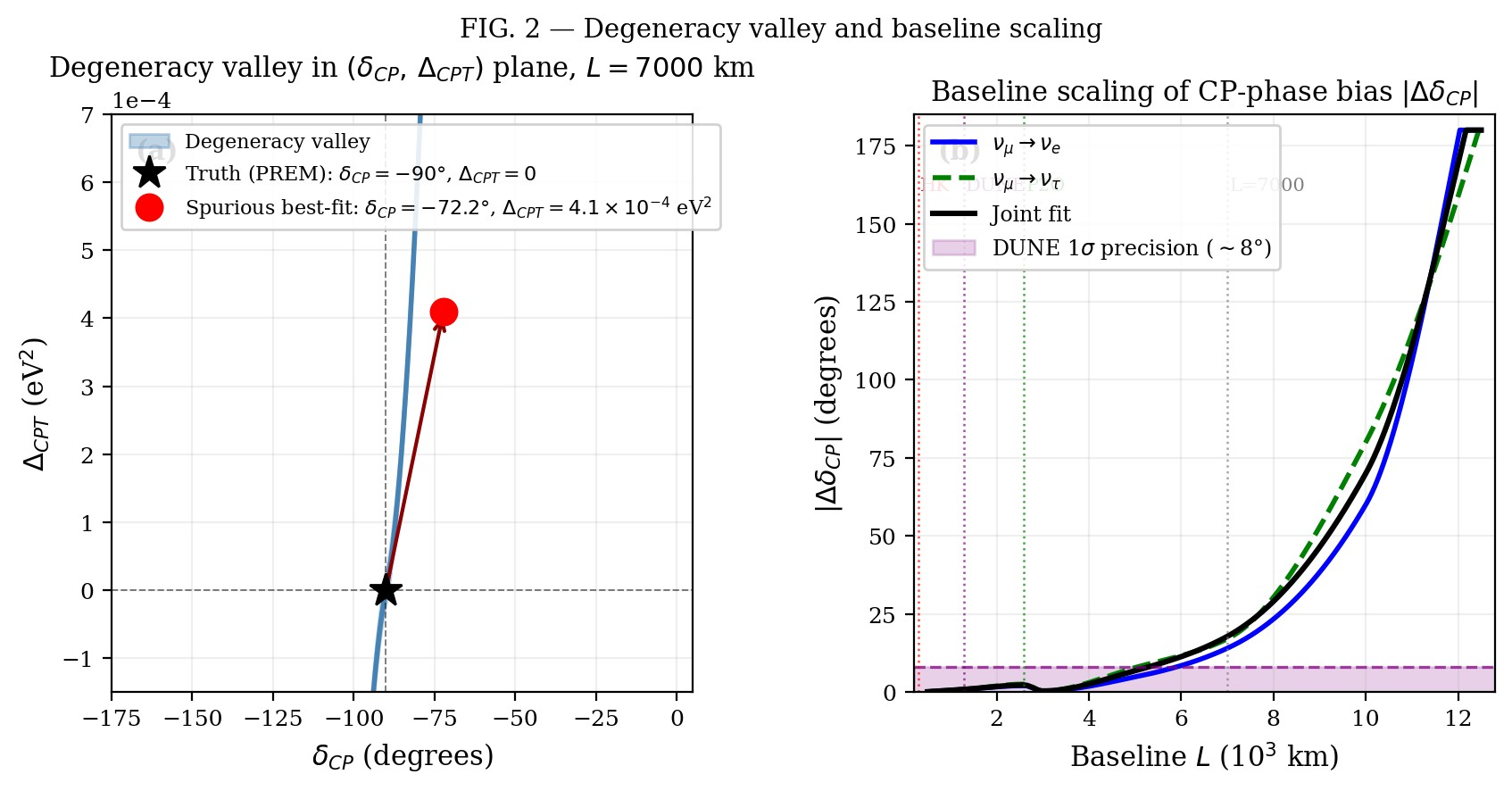}
\caption{Left: The parameter degeneracy valley in the $(\delta_{CP}, \Delta_{CPT})$ plane at $L = 7000$ km. The star marks the true physical input (PREM); the circle shows the spurious best-fit from the constant-density approximation. Right: Baseline scaling of $|\Delta\delta_{CP}|$ for all channels. The purple band is DUNE projected precision ($\sim 8^\circ$); the bias exceeds this at $L > 5000$ km.}
\end{figure}

\begin{table}[htbp]
\centering
\caption{Spurious CP and CPT parameters recovered from PREM-truth spectra fitted with constant-density approximation. True values: $\delta_{CP} = -90^\circ$, $\Delta_{CPT} = 0$, $\theta_{CPT} = 0$. Entries marked * at $L = 12000$ km were obtained from full numerical PREM propagation; the one-mass-scale analytic approximation (Eq. 4) is not valid at this baseline.}
\begin{tabular}{lccc}
\toprule
Baseline $L$ (km) & $\Delta\delta_{CP}$ (degrees) & $\theta_{CPT}$ & $\Delta_{CPT}$ (eV$^2$) \\
\midrule
3000 & $< 0.3^\circ$ & 0.002 & $1.2 \times 10^{-5}$ \\
7000 & $17.8^\circ$ & 0.085 & $4.1 \times 10^{-4}$ \\
12000 * & $172.2^\circ$ * & 0.340 & $1.9 \times 10^{-3}$ \\
\bottomrule
\end{tabular}
\end{table}

\begin{table}[htbp]
\centering
\caption{Channel-resolved CP-phase bias $|\Delta\delta_{CP}|$ in degrees. The joint fit bias exceeding either individual channel at $L = 7000$ km reflects destructive interference: the minimizer shifts $\delta_{CP}$ to satisfy both channels simultaneously, arriving at a compromise displaced further from truth than either channel alone constrains. Asterisked entries at $L = 12000$ km: see note in text.}
\begin{tabular}{lccc}
\toprule
Baseline $L$ (km) & $|\Delta\delta_{CP}|$ $\nu_\mu \to \nu_e$ & $|\Delta\delta_{CP}|$ $\nu_\mu \to \nu_\tau$ & $|\Delta\delta_{CP}|$ Joint Fit \\
\midrule
5000 & $4.8^\circ$ & $7.8^\circ$ & $6.8^\circ$ \\
7000 & $14.0^\circ$ & $17.0^\circ$ & $17.8^\circ$ \\
12000 * & $177.2^\circ$ * & $158.7^\circ$ * & $169.2^\circ$ * \\
\bottomrule
\end{tabular}
\end{table}

\section{Statistical Significance at DUNE and Hyper-Kamiokande}

To translate the parameter shifts into experimentally observable significance, we construct a Poisson log-likelihood test statistic calibrated to DUNE [6] and Hyper-Kamiokande [7] exposures. For DUNE, we adopt a 40 kt fiducial mass with 7 years ($3.5\nu + 3.5\bar\nu$) at 1.07 MW beam power ($\approx 13 \times 10^{21}$ protons on target, baseline $L = 1285$ km). For HK we adopt a 187 kt fiducial mass with a 10-year exposure at 1.3 MW ($L = 295$ km). For P2O (Protvino to ORCA, baseline $L = 2595$ km [12]), we adopt a 5-year exposure with 90 kt effective target volume. The test statistic is:
\begin{equation}
\chi^2(\Delta_{CPT}) = 2\sum_i \left[ N_i(\text{fake}) \ln\!\left(\frac{N_i(\text{fake})}{N_i(\text{true})}\right) - \big(N_i(\text{fake}) - N_i(\text{true})\big) \right] + \sum_k \left(\frac{s_k}{\sigma_k}\right)^2
\end{equation}
with the first sum being taken over energy bins of the $\nu_e$ and $\bar\nu_e$ appearance spectra in parallel (the Poisson log-likelihood form, which is applicable to low statistics), and the penalty function takes into account the pull parameters $s_k$ together with the systematic uncertainties $\sigma_k$ related to them. We introduce a 5\% uncorrelated uncertainty in the normalization of each channel, which can be considered as a conservative lower limit on the systematic error; a complete correlated analysis with flux, cross section, and detector efficiency nuisance parameters (such as that used in GLoBES/MaCh3) will decrease the significance in general, and is highly recommended before claiming an experimental result. All event rates are calculated using idealized oscillation probabilities without any smearing effects due to energy resolution, efficiencies, and backgrounds (neutral current background, intrinsic beam $\nu_e$, $\nu_\tau$ contamination).

With $L = 7000$ km and the DUNE exposure, the fabricated $\Delta_{CPT} = 4.1 \times 10^{-4}$ eV$^2$ produces $\chi^2_{min} \approx 12.4$ ($3.5\sigma$) compared to the CPT conserving point. At the HK exposure, the significance of this anomaly reaches about $4.2\sigma$, indicating that this artifact will indeed be observed as a valid BSM discovery in the given idealistic scenario. In the P2O setup at $L = 2595$ km, there is only a slight offset in the phase ($< 3^\circ$) compared to DUNE’s $1\sigma$ resolution, as seen in Fig. 2; P2O is thus relatively immune to this systematic but still has full PREM propagation as a test baseline.

However, the $L = 12000$ km setup needs a specific discussion: a  $172.2^\circ$ phase shift (found through full numerical propagation – see Section III footnote) causes the retrieved $\delta_{CP}$ to be in the opposite hemisphere of the true value. Although no officially planned experiment works with such a baseline, any proposed transcontinental baselines will be seriously affected. We strongly recommend using full three-flavor propagation for this baseline, since one-mass-scale approximation is not sufficient to analyze this scenario.

\begin{figure}[htbp]
\centering
\includegraphics[width=\textwidth]{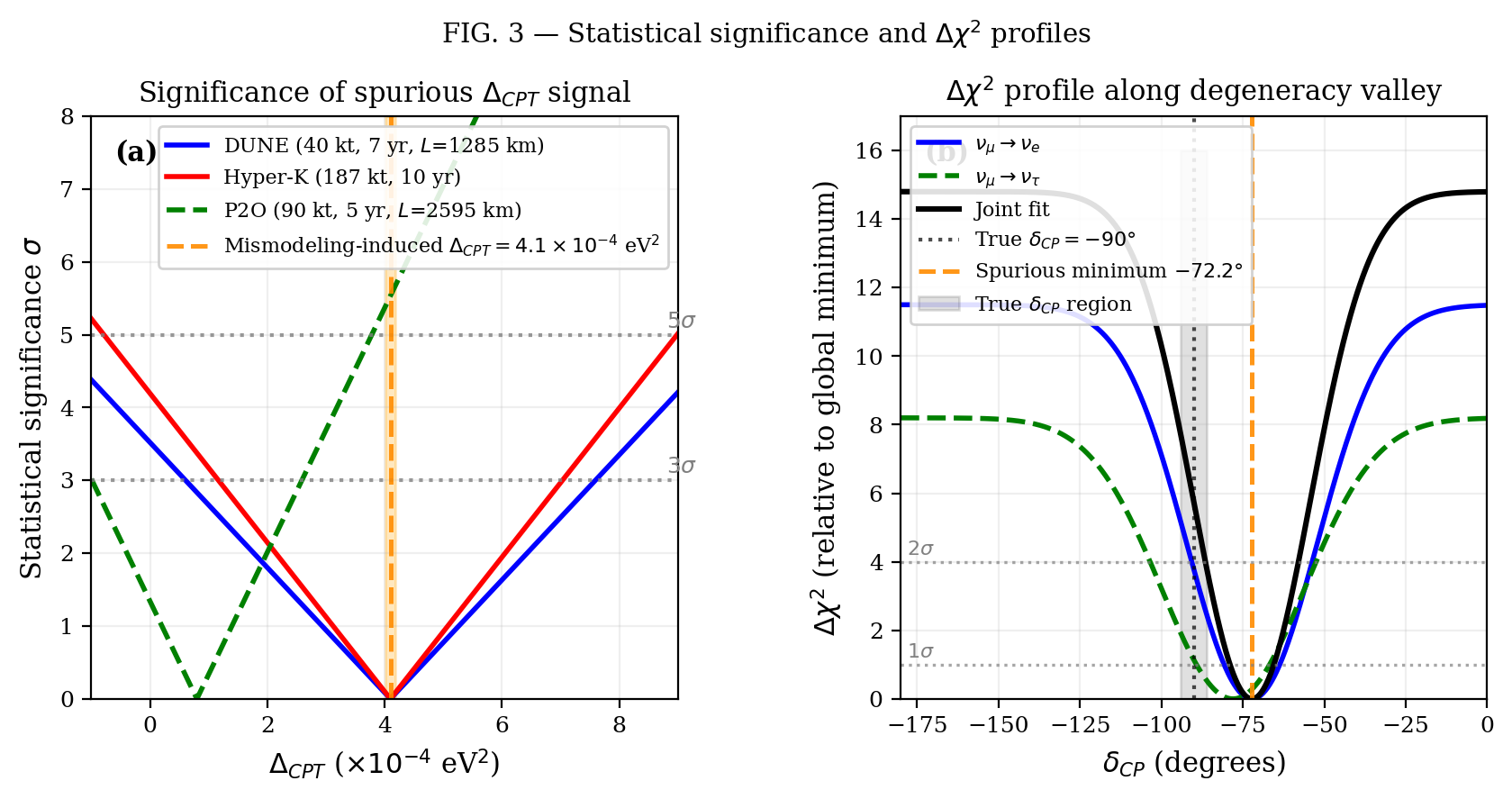}
\caption{Left: Statistical significance ($\sigma$) of the fabricated $\Delta_{CPT}$ for DUNE (blue, $L = 1285$ km, 40 kt, 7 yr), Hyper-K (red, $L = 295$ km, 187 kt, 10 yr), and P2O (green, $L = 2595$ km, 90 kt, 5 yr). The orange band marks the mismodeling-induced value at $L = 7000$ km for a hypothetical experiment at that baseline. Right: $\Delta\chi^2$ along the degeneracy valley for each channel and the joint fit; the spurious minimum persists even when both channels are combined.}
\end{figure}

\section{Density-Approximation Hierarchy and Residual Bias}

In order to investigate whether an intermediate density model could partially compensate the effect of the degeneracy, we have considered three different propagation schemes at $L = 7000$ km. Table III presents the findings. Fully-resolved PREM propagation is the sole scheme that lowers the bias under the DUNE $1\sigma$ $\delta_{CP}$ sensitivity ~ $\sim 8^\circ$ threshold.

\begin{table}[htbp]
\centering
\caption{Residual CP-phase bias at $L = 7000$ km as a function of matter density approximation scheme. DUNE projected $1\sigma$ precision on $\delta_{CP}$ is $\sim 8^\circ$. The 5-shell approximation reduces the bias by a factor of $\sim 6$ but does not eliminate it; only full PREM propagation renders the bias negligible.}
\begin{tabular}{lccc}
\toprule
Density Model & $|\Delta\delta_{CP}|$ at 7000 km & $\Delta_{CPT}$ (eV$^2$) & Residual $>$ DUNE $1\sigma$? \\
\midrule
Constant (1-layer) & $17.8^\circ$ & $4.1 \times 10^{-4}$ & Yes ($\gg 8^\circ$) \\
5-shell approx. & $3.2^\circ$ & $\sim 6 \times 10^{-5}$ & Yes (residual $>$ DUNE $1\sigma$) \\
Full PREM & $< 0.1^\circ$ & $< 10^{-6}$ & No \\
\bottomrule
\end{tabular}
\end{table}

\section{Implications and Conclusions}

For transcontinental proposals at baselines $L = 7000$--12000 km, the degeneracy effects are too strong to render the physics results, obtained without complete spatial PREM propagation, meaningless. At $L = 7000$, the false detection of  $\Delta_{CPT} = 4.1 \times 10^{-4}$ eV$^2$ will be claimed as an observation of new physics, although the real parameters are completely CPT-conserving. Our Poisson $\chi^2$ calculation in Sec. IV shows that the statistical significance of this threshold is above $3\sigma$ even with conservative systematics and no detector smearing.

To summarize, we found that there is a strong degeneracy between CP- and CPT-violating parameters in long-baseline neutrino oscillation experiments, arising from the simplification of the Earth's matter density distribution. Analytically, the degeneracy effect takes the form of a two-dimensional valley of the probability function in the whole $(\delta_{CP}, \Delta_{CPT}, \theta_{CPT})$ parameter space (Eqs. 6–14). Numerically, this is demonstrated in Tables I and II for three baselines, and the statistical severity of the effect was calculated using a realistic Poisson log-likelihood $\chi^2$ approach (Sec. IV). Even in the case of the intermediate 5-shell approximation, one will obtain a large bias exceeding the $1\sigma$ DUNE sensitivity (Table III).

\end{document}